\documentclass[11pt]{amsart}
\usepackage{kept}
\usepackage{graphicx}
\begin{document}

\title[jSET - The Java Software Evolution Tracker]{Java Software Evolution Tracker}

\author[Arthur-Jozsef Molnar]{Arthur-Jozsef Molnar$^{(1)}$}
\address{$^{(1)}$ Department of Computer Science, Faculty of Mathematics and Computer Science,Babe\c{s}-Bolyai University, 1, M. Kogalniceanu, Cluj-Napoca 400084, Romania}
\email{arthur@cs.ubbcluj.ro}

\date{31.03.2011} % submission date

% AMS Subject Classification (2010)

% ACM Computing Classification System (1998)
% I think there' an error here (no studia package used)
%\subjclassCR{
%D.2.2 [\textbf{Software}]: Software Engineering -- \textit{Design Tools and Techniques}
%}

\begin{abstract}
This paper introduces the Java Software Evolution Tracker, a visualization and analysis tool that provides practitioners the means to examine the evolution of a software system from a top to bottom perspective, starting with changes in the graphical user interface all the way to source code modifications.
\end{abstract}

\maketitle

\section{Introduction}
Software tools occupy an important place in every practitioner's toolbox. They can assist in virtually all activities undertaken during the life of software starting from requirements analysis to test case design and execution. By studying the evolution of widely used IDE's such as Eclipse \cite{6,17} one can see that each new version ships with better and more complex tools for aiding professionals in building higher quality software faster. Modern environments include tools for working with UML artifacts, navigating source code and working with a wide variety of file types.

However, modern day software systems fall into many categories, each having unique requirements, artifacts and processes. Recent hardware advances enabled new devices with large screens running rich user interfaces. Unfortunately, while this trend is in full swing, the same cannot be claimed about the state of the tools that should support it. A look at today's software tools reveals that while most do enable some visualizations there is a clear lack of advanced tools enabling unified program visualisation and analysis from GUI layer right into the source code.  

As such, our goal is to apply the latest achievements in research in the development of new, useful tools for practitioners looking to build GUI-based software. The jSET application was developed as a first step in the direction of integrating domain-specific knowledge and academic research results into useful applications for software practitioners. The main advantage of jSET is that by using state of the art tools from the academic community, it enables new visualizations that unify the GUI with the application code in a unitary whole. Given software's evolutionary nature, jSET allows visualizing how the target application changes across versions, providing support for tracking the software's evolution.

The rest of this paper is structured as follows: the next section introduces the work jSET is based on. The third section describes the tool in detail, while the fourth overviews its current limitations. The last section is reserved for conclusions and future work planned.

\section{Related work}
The development of jSET was made possible by two tools that come from the academic environment. They are presented in the following paragraphs together with earlier efforts of using them for software visualisation.

The first of the employed tools is called GUIRipper and is part of the comprehensive GUITAR toolset \cite{5}. The GUIRipper acts on a GUI driven target application \cite{8} that it runs and records all the widgets' properties across all the application's windows. It does this by starting the target application, recording the properties of all the widgets created on the application's starting windows and firing events on them (e.g: clicking buttons) with the purpose of opening the application's other windows that are then recorded in turn. The resulting GUI model, described in detail in \cite{8}, is persisted in XML format for later use. It is important to note that the only required artifact is the target application's compiled code (or bytecode for a Java application). Although completely automated, GUIRipper's behaviour can be customized by configuration files. This makes it possible to avoid firing events with unwanted results, such as creating network connections, printing documents and so on. The GUIRipper tool is available in versions that work with Microsoft Windows and Java applications \cite{12}. The jSET tool uses the Java implementation of GUIRipper\footnote{Called JFCGUIRipper}. When developing jSET, additional functionality for recording widget event handlers and capturing screenshots was programmed into GUIRipper. This modified version can be found on the jSET website \cite{9}.

The second application is the Soot analysis framework \cite{1,2,11}. Soot is a static analysis framework targeting Java bytecode; all its implemented analyses are performed without running the target application. Currently there are many types of analyses Soot can perform \cite{2,11}, some of which are planned for future integration with our tool. One of the most important artifacts produced by Soot is the application's call graph: a directed graph that describes the calling relations between the target application's methods \cite{2}. The graph's vertices represent methods while the edges model the calling relations between them. Being computed statically, it does not provide information regarding the order methods are called or execution traces. This static callgraph is an over-approximation of all the dynamic callgraphs obtained by running the application on all its possible inputs. Of course, this means the graph will contain spurious edges and a number of algorithms were devised to reduce their number. The interested reader is referred to \cite{11} for a detalied comparison of the implemented algorithms. By default, the Soot wrapper implemented for jSET uses the algorithm detailed in \cite{2}, which provides a very good approximation of the application callgraph \cite{11}. It is important to note that since all non-trivial Java applications call methods within the platform, most of them also using third party libraries, they all must be incorporated in the call graph. This usually leads to a complex structure that is intrinsically difficult to visualize without abstracting away some of the data \cite{4}. The abstractions implemented in jSET for compacting this spurious data are detailed in the following section.

The applications described above laid the groundwork for the development of advanced software tools. Some of these earlier efforts that served as inspiration for jSET's development are discussed in the following paragraphs.

Possibly the earliest of such tools is JAnalyzer \cite{3}, \emph{a visual static analyzer for Java} developed by Bodden et al. JAnalyzer leverages the call graph information generated by Soot and graphically displays the calling relations in a program. It also implements a Java source code parser that allows viewing the source code for application methods, thus providing a link between the bytecode and its sources.

A more advanced approach was undertaken in \cite{4} where the author presents a call graph comparison tool that ranks differences according to their importance. The same paper also introduces a browser application for navigating call graphs, similar to JAnalyzer.

An interesting approach to software visualization in a language independent manner was proposed by Rajala et. al \cite{26} in the form of VILLE. Although built for didactic use, VILLE proposes some interesting ideas like support for multiple languages, execution tracing and call stack visualization.

More recent approaches have attempted to enrich IDE software with visualization capabilities. One of these approaches is Code Bubbles, developed by Bragdon et al. \cite{27}. Code Bubbles proposes a unitary view of a program's sources increasing developer productivity and minimizing overhead. Altough not a software visualizer per se, Code Bubbles proposes a tight integration of visualization tools with modern IDEs for maximum efficiency. Building on this effort, Microsoft Research integrated Code Canvas \cite{28} into Visual Studio 2010. Code Canvas provides a unified view of the source code together with all related information for easy synthetization of information.

For the interested reader, a detailed evaluation concerning software visualizers that takes into account effectiveness and presentation techniques is avalable in \cite{29}.

\section{jSET - Java Software Evolution Tracker}
jSET is an analysis and visualisation tool created for software practitioners and researchers alike. The main ideas guiding its development are:
\begin{enumerate}
\item Provide advanced visualization tools for easily accessing static analysis results inside a software project environment without the need for a laborious setup phase.
\item Integrate the obtained results in the context of GUI driven applications by offering seamless transition in visualization from GUI level down to viewing the application's source code.
\item Facilitate identification and analysis of changes across versions of a software system from GUI changes to source code modifications.
\end{enumerate}
In order to use jSET, the first step is to create a project. A jSET project is an XML file that contains information about the locations off all the necessary artifacts. Its purpose is to capture the state of the target application at a given moment in time. In order to have a valid project, the following data is required:
\begin{itemize}
\item The GUI model obtained by running our modified version of GUIRipper on the target application.
\item The file containing the application's callgraph obtained by running our Soot wrapper on the target application.
\item The target application's bytecode (including used libraries).
\item The target application's source code \footnote{Only if viewing the source code is desired}.
\end{itemize}
It is important to note that all the steps of building a project can be easily automated. Both GUIRipper and our Soot wrapper can be executed via command line and manual intervention using their configuration files is required only for certain changes in the target application such as specifying special handling for some GUI elements (e.g: exempting components from analysis) or updating the application's libraries. This approach makes jSET easily integrateable into the target application's build system. The jSET website \cite{9} is home to a collection of projects that track the evolution of two widely used open source projects \footnote{FreeMind - http://freemind.sourceforge.net}\footnote{jEdit - http://sourceforge.net/projects/jedit}. This repository includes all the target application code and scripts used for building the projects.

\begin{figure}[htbp]
	\centering
	\includegraphics[width=\textwidth]{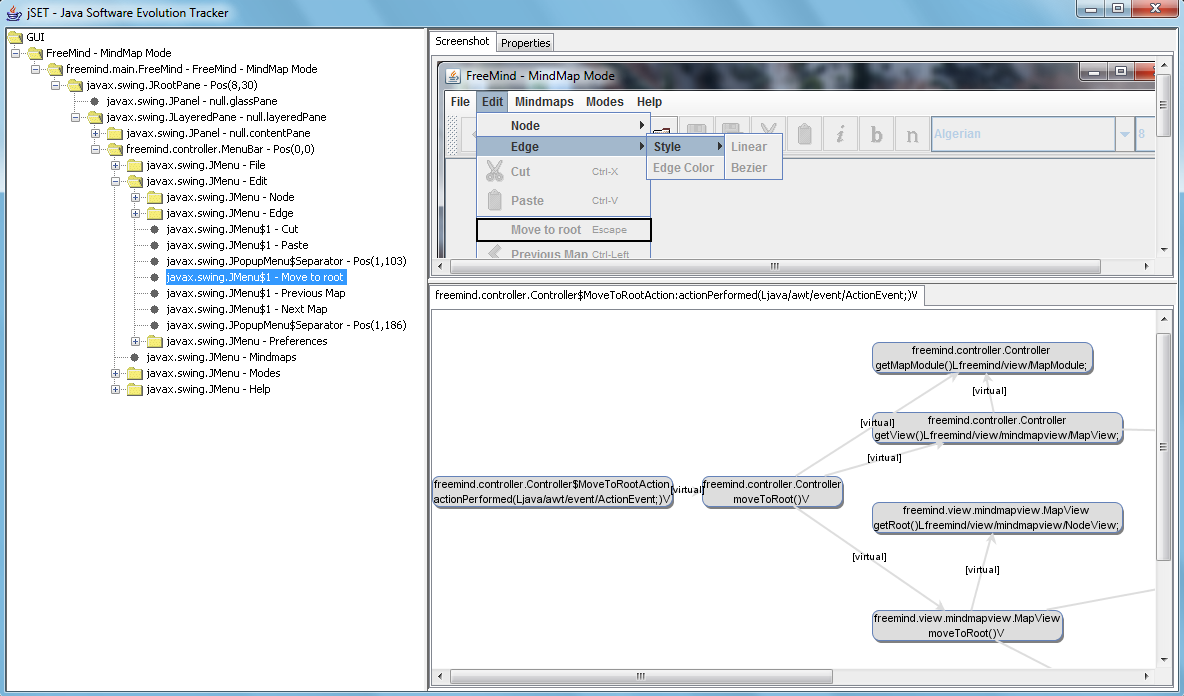}
	\caption{jSET in Project Exploration mode}
	\label{jSET1}
\end{figure}

The jSET application can be used in two modes: project exploration and project comparison. When starting the application, the user must select one or two projects to load. Selecting one will default jSET to the project exploration mode. When started in comparison mode, jSET can be used to display the differences between the target application's versions\footnote{The projects \emph{should} capture the same application at different versions, however this is not enforced}. Figure \ref{jSET1} shows jSET in exploration mode, while Figure \ref{jSET2} shows the tool's comparison mode. In both screenshots, the target application is an early version of the open-source FreeMind software. The tool's user interface is rather similar for both modes but because of differences between displayed information, the following paragraphs will present them in detail, starting with the simpler mode of project exploration.

\subsection{Project exploration mode.} As stated before, this mode can be started by selecting a single project when the tool is started. jSET's user interface consists of several panes displaying information about the loaded project. The left hand side pane displays the user interface hierarchy of the target application, as captured by GUIRipper. When a GUI element is selected from the hierarchy, the tabbed pane on the right hand side of Figure \ref{jSET1} will display the selected widget's properties\footnote{Like swingExplorer (www.swingexplorer.com)} and a screenshot of the target application, taken by GUIRipper with the selected widget highlighted. 

Among the displayed properties we can find the handlers associated with the widget's events (listeners in Java terminology). Some of these are attached by the platform itself as they control behind the scenes aspects regarding the GUI. Other event handlers are defined by the application itself. One of jSET's original contributions concerns the visualisation of the target application's event handling. When selecting one of these handlers, the relevant part of the application's call graph is displayed in the right lower pane, as seen in Figure \ref{jSET1}. Here it is possible to examine what methods might be run when a certain event is fired (e.g: a button is clicked on the GUI).

\begin{figure}[htbp]
	\centering
	\includegraphics[width=\textwidth]{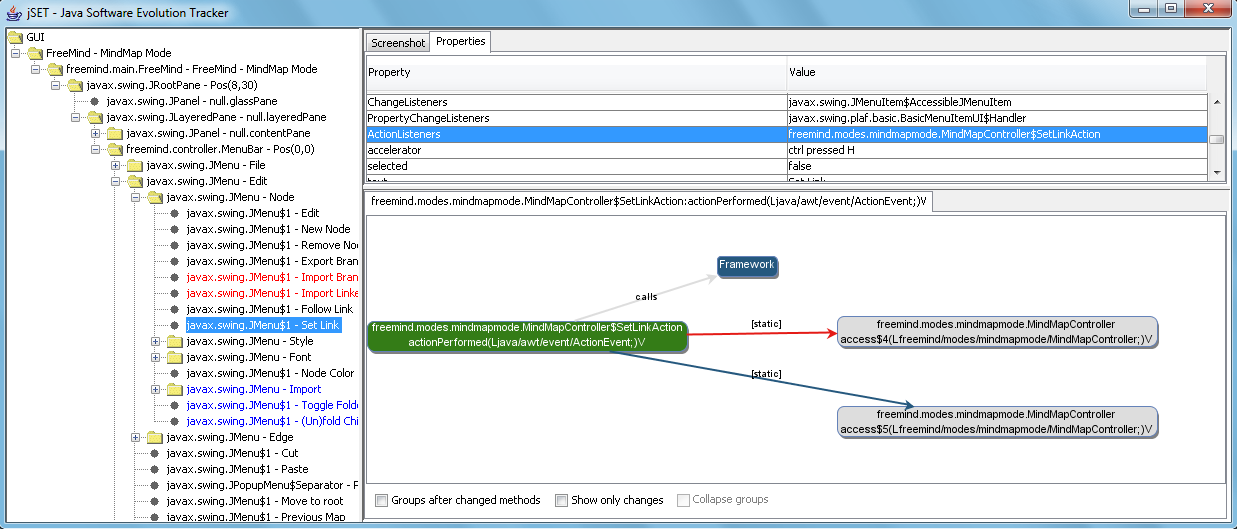}
	\caption{jSET in Project Comparison mode}
	\label{jSET2}
\end{figure}

The previous section discussed the inherent complex nature of a statically built callgraph. From the author's experience, backed by empirical research detailed in \cite{4,18} most of the methods in a callgraph will belong to the Java platform itself. Since we are interested in analyzing calling relations within application code, our Soot wrapper categorizes all methods in the callgraph as framework, library\footnote{Usually found on the application's classpath} or application methods. Non-application methods are abstracted in the call graph display pane by nodes labeled as \emph{``Framework''}, as seen in Figure \ref{jSET2}. Application methods that call framework or library code will have edges to these nodes. 

As a result, nodes explicitely represented are the event's handler method and all the application methods that it might call transitively. A future direction of development worth mentioning regards allowing more information to be obtained for framework and library calls including possible callbacks into application code without burdaining users with large amounts of superfluous information.

Visualizing the target application's call graph has limited value if the source code behind it cannot be easily consulted. Therefore, jSET uses the Eclipse Java development tools \cite{13} that include a Java source parser for building the abstract syntax tree of the provided source files. It can thus match compiled methods with their Java sources allowing users to browse the source code for methods shown in the graph display pane. Right clicking a displayed method brings up a menu with options for displaying its source or bytecode. It is important to note that while currently our tool only supports viewing Java source code, compatible bytecode can be compiled from other languages such as Haskel, Eiffel or Ada \cite{19}. The bytecode can of course be consulted for all the methods displayed, regardless of source language.

\subsection{Project comparison mode.} As stated above, the compare view's layout (in Figure \ref{jSET2}) is similar to the project exploration one, so this section will only discuss the major differences between the modes. 

The first such difference regards the GUI tree shown on the left hand side. While the exploration mode displayed the target application's GUI hierarchy, the comparison mode also displays the differences between the two project's GUIs. Looking at Figure \ref{jSET2}, we can see certain items from the hierarchy are color coded. Red items represent widgets that can no longer be found on the newer version, while items in blue are widgets that were not found on the older one. Green items represent widgets affected by underlying changes in their event handler code. This hierarchy is computed by comparing the hierarchies of the loaded projects; GUI elements are matched by their extracted properties. Unfortunately, the current implementation for GUI element matching is not without pitfalls, as it is sensitive to changes in the structure of the user interface. Resizing, moving or changing GUI elements' places in the hierarchy may lead to them not being correctly recognized, causing them to appear as duplicates in the final hierarchy. Research regarding identification of equivalent GUI components across versions is an open problem. Early work \cite{7} reported encouraging results and we consider the jSET tool to be a good platform for more advanced research on the topic. 

The second difference between jSET's compare and exploration modes regards the graph display. For those components that cannot be matched across versions (represented by red and blue in the GUI hierarchy) the partial call graph displayed will be the same as in exploration mode. However, for widgets identified in both versions, a new call graphs visualization was developed as shown in Figure \ref{jSET2}. The astute reader will notice the same color coding used as with the GUI hierarchy, this time customized for application calls. As such, the displayed graph will actually be the reunion of the event handlers' call graphs across versions. Red edges represent call relations removed from the newer version while blue edges show new calling relations. Green nodes represent methods that underwent changes in their code, while unchanged methods remain light gray. 

Even so, for complex application methods the displayed section of the callgraph might contain too many nodes to be easily browsable. jSET addresses this issue via the toolbar at the bottom of the graph display. It contains controls that allow unmodified method nodes\footnote{These are generally the \emph{uninteresting} ones in project comparison} to be grouped in collapsed subgraphs, leaving only methods that were changed in plain view. Empirically we observed this approach to solve the great majority of cases where the displayed graph was deemed too complex.

In addition to the exploration mode, right clicking a changed method node (in green) will bring up a menu allowing the source code of methods to be compared across versions using an implementation of the diff algorithm \cite{14}. This enables jSET to trace back to a compiled method's sources, enabling users to view or compare the source code between application versions.

The jSET application is an ongoing effort of providing useful tools based on the latest accomplishments in research. Its exploration mode provides an integrated view of an application linking the easily browsable GUI to the source code behind it. To our knowledge, jSET is the first application to accomplish this for generic Java software. This mode is useful for understanding how the target application works by identifying events that cause code to run and providing visualizations for the calling relations between application methods. It can help people unfamiliar with the target application in learning about its event handling and observing the link between the application's GUI and its sources. This mode is also useful for checking which GUI element might cause a certain method or method chain to be called, helping with maintaining good application design.

However jSET's most important contribution is its comparison mode. While modern software IDE's provide advanced tools for tracking source file changes, jSET improves this by providing application level change visualizations: the evolution of the user interface, calling relations and source code can be traced using the provided visualizations.

Testers can use this mode to determine what areas of the GUI are newly implemented or have been recently changed and adjust the testing plans accordingly. The GUI tree and call graph visualizations also provide valuable information about the unchanged areas of the application that do not need regression testing. The tool also enables users to easily assess the magnitude of changes across versions and on a broader scale to track the evolution of the target application across multiple versions.

\section{Limitations}
Although much thought went into the design and implementation of jSET, there are some aspects that limit its usability. Some of these stem from inherent limitations of the tools jSET itself is based on. The following list attempts an overview of these limitations:
\begin{itemize}
\item \emph{Dynamic user interfaces.} While the GUIRipper can be considered a mature tool, it is not capable of fully recording every application's GUI. Some applications create and dispose of GUI elements dynamically; recording these would require using a descriptive language for specifying rules that govern GUI element creation and disposal, a task bringing added complexity to the process. Event handlers added or removed during program execution might be missed by GUIRipper leading to an incompletely recorded GUI and affecting the accuracy of visualizations. Also, user interfaces that have timing issues (e.g: web interfaces) or that present a continuous stream of data (e.g: media players) cannot be completely captured by the tool \cite{8}.
\item \emph{Native methods.} Soot's algorithms for call graph building work on Java bytecode. Java code however can call native methods \cite{10} that cannot be analyzed.\cite{1} mitigates this by manually overviewing the effects of the native code called. However, if a virtual machine that has not been pre-analyzed is used or the application itself calls native methods, the obtained call graph might be incomplete.
\item \emph{Reflection.} Applications using reflection can instantiate classes and call methods that the call graph does not include by default. For these situations, Soot can be given a list of classes that can be instantiated by reflection \cite{11} to incorporate in the call graph. Since this is a manual undertaking, it might prove time consuming and is error prone.
\item \emph{Interacting widgets.} In some cases, events fired on widgets might create new events on other GUI elements (e.g: AbstractButton's doClick() ). This is not accounted for by the current version of jSET, and in these cases library callbacks might occur that are not captured in the displayed callgraph.
\end{itemize}

\section{Conclusions and future work}
In this paper we presented jSET, a new software visualisation and analysis tool. jSET introduces a new top to bottom approach for software visualization starting at the GUI level and ending at the source code itself. 

At a high level, we showed a new way of identifying and displaying changes in the target application's GUI across versions. We also showed a new way of examining the target application's source code by starting from events fired by the GUI. Interprocedural analysis generated by Soot was harnessed in developing a compact view to compare calling relations between application versions.

We believe jSET is a useful tool for software practitioners. However there are many ways in which its functionality can be further improved. A direction of research integrateable into present efforts regards creating new algorithms for matching GUI elements across application versions. Initial work on the topic \cite{7} shows promising results and we believe the inclusion of code analysis can bring further improvements on the state of the art.

Another direction of work regards tracking a target application's evolution across multiple versions. As we have shown, jSET is able to provide compare views for distinct versions of an application. It is our desire to generalize this approach in order to enable viewing evolution across more than two versions in a single jSET instance. This would allow fine-grained, incremental visualizations for assessing the evolution of an application.

Of course, important efforts must be dispensed regarding the current limitations of the tool; code analyzed via Soot could be used to ascertain interconnected events so library callbacks can be displayed whenever they might occur \cite{16}. A mechanism for detected and unresolved uses of reflection should be reported so that it can also be taken into account.

A more elaborate direction of research concerns integrating visualizations provided by jSET with artifacts specific to model driven architecture approaches, in both desktop \cite{22} and web based \cite{20,22} applications. This would broaden our tool's scope as a software visualizer enabling its use in a wider variety of contexts. While such an extension requires additional research, the available literature reports promising results \cite{23,24} regarding the application of static analyses to web-based software.

A more distant idea is using jSET as the visualisation platform of an automated regression testing procedure for GUI based applications \cite{25}. Having the means of visualizing key application information across versions, jSET could be used for visualising test results, guiding test suite generation, test data input and automated test execution.

\section*{Acknowledgements}
The author was supported by programs co-financed by The Sectoral Operational Programme Human Resources Development, Contract POS DRU 6/1.5/S/3 - ``Doctoral studies: through science towards society''

\label{finalpage}

\bibliography{biblio}
\bibliographystyle{plain}

\end{document}